\documentstyle[psfig]{mn}
\def\lesssim{\mathrel{\hbox{\rlap{\hbox{\lower4pt\hbox{$\sim$}}}\hbox{$<$}}}}
\def\gtrsim{\mathrel{\hbox{\rlap{\hbox{\lower4pt\hbox{$\sim$}}}\hbox{$>$}}}}
\def\la{\mathrel{\hbox{\rlap{\hbox{\lower4pt\hbox{$\sim$}}}\hbox{$<$}}}}
\def\ga{\mathrel{\hbox{\rlap{\hbox{\lower4pt\hbox{$\sim$}}}\hbox{$>$}}}}

\def\aap{A\&A}
\def\aaps{A\&A}

\def\aj{AJ}

\def\apj{ApJ}

\def\apjl{ApJL}

\def\araa{ARA\&A}

\def\mnras{MNRAS}

\def\ie{{\it i.e.\ }}

\def\spose#1{\hbox to 0pt{#1\hss}}
\def\approxlt{\mathrel{\spose{\lower 3pt\hbox{$\sim$}}
	\raise 2.0pt\hbox{$<$}}}
\def\approxgt{\mathrel{\spose{\lower 3pt\hbox{$\sim$}}
	\raise 2.0pt\hbox{$>$}}}

\def\<{\thinspace}
\def\boxit#1{\vbox{\hrule\hbox{\vrule\kern3pt\vbox{\kern3pt
          #1 \kern3pt}\kern3pt\vrule}\hrule}}

\def\ga{{\rm\thinspace gauss}}

\def\msun{\hbox{$\rm\thinspace M_{\odot}$}}

\def\yr{{\rm\thinspace yr}}
\def\h50{\hbox{$\rm\thinspace h_{50}$}}
\def\h50m1{\hbox{$\rm\thinspace h_{50}^{-1}$}}

\begin{document}
\title[The Evolution of Stellar Exponential Discs]{The Evolution of
Stellar Exponential Discs} \author[A. M. N. Ferguson \& C. J.
Clarke]{A. M. N. Ferguson\thanks{Present address:  Kapteyn Astronomical
Institute, Postbus 800, 9700 AV Groningen, The Netherlands} \& C. J.
Clarke \\ Institute of Astronomy, University of Cambridge, Madingley
Road, Cambridge CB3 0HA}
\date{\today}
\maketitle

\begin{abstract}

Models of disc galaxies which invoke viscosity-driven radial flows have
long been known to provide a natural explanation for the origin of
stellar exponential discs, under the assumption that the star formation
and viscous timescales are comparable.  We present models which invoke
simultaneous star formation, viscous redistribution of gas and
cosmologically-motivated gaseous infall and explore the predictions
such models make for the scale length evolution and radial star
formation history of galactic stellar discs.  While the inclusion of
viscous flows is essential for ensuring that the stellar disc is always
exponential over a significant range in radius, we find that such
flows play essentially no role in determining the evolution of the disc
scale length.  In models in which the main infall phase precedes the
onset of star formation and viscous evolution, we find the exponential
scale length  to be rather invariant with time, with the bulk of the
disc stars at all radii out to $\sim5$ scale lengths being relatively
old (ie. ages $\gtrsim 6-8$~Gyr for an assumed disc age of 11~Gyr).
On the other hand, models in which star formation/viscous evolution and
infall occur concurrently result in a smoothly increasing scale length
with time, reflecting the mean angular momentum of material which has
fallen in at any given epoch. The disc stellar populations in these
models are predominantly young (ie. ages $\lesssim 5$~Gyr) beyond a few
scale lengths.  In both cases, viscous flows are entirely responsible
for transporting material to very large radii.  Our results are robust
for a range of currently popular star formation laws and  infall
prescriptions.  We discuss existing observational constraints on these
models from studies of both local and moderate redshift disc galaxies.
In particular, a  good agreement is found between the solar neighbourhood
star formation history predicted by our infall model and the recent
observational determination of this quantity by Rocha-Pinto et al (2000).
\end{abstract}

\begin{keywords}
galaxies: formation - galaxies: evolution - galaxies: structure -
galaxies: spiral - galaxies: fundamental parameters
\end{keywords}

\section{Introduction}
\label{sec:intro}
One of the most notable properties of disc galaxies is that the surface
brightness profiles of their stellar discs are remarkably close to
exponential distributions over a large dynamic range in radii (eg. de
Vaucouleurs 1959; Freeman 1970; de Jong 1996a).  This has traditionally
been held to imply an exponential surface density profile for gas
infalling to the galactic plane, which in turn places particular
constraints on the detailed angular momentum distribution of baryonic
material in the protogalaxy.   Gunn (1982), for example, showed that,
under the assumption of angular momentum conservation for each gas
element under infall, a uniformly rotating uniform sphere collapses to
an approximately exponential profile, however the resulting function is
only exponential over $\sim 3$ scale lengths compared to the more
impressive 4-6 scale lengths exhibited by real galaxies (eg. de Jong
1996a).  Calculations with other initial halo density profiles and
angular momentum distributions produce similar results (eg. van der Kruit 1987;
Dalcanton, Spergel \& Summers 1997; Bullock et al 2000).

The exponential profile is so ubiquitous, and its appearance when
plotted as magnitude versus radius so deceptively simple, that one may
be tempted to regard it as a natural or obvious outcome of
gravitational collapse.  In fact, gravitational collapse calculations
in other astronomical contexts never give rise to exponential disc
profiles (Cassen \& Moosman 1981; Terebey, Shu \& Cassen 1984), with
the more usual outcome being profiles of approximately power law form.
In the galactic context, the difficulty of producing exponential
profiles over a large number of scale lengths through angular momentum
conserving collapse from reasonable initial conditions has recently
been highlighted by Efstathiou (2000): the outer regions of galactic
discs contain so little mass, that for power law halo density profiles,
the outer disc material must originate from a small range of radii in
the halo. On the other hand, these regions contain a large dynamic
range of angular momentum, so that these conditions can only be met
simultaneously if the angular momentum profile at the halo suffers an
abrupt, and physically implausible, upturn at some radius (see Figure
6(b) of Efstathiou 2000).

Exponential profiles have the unique feature that the scale length on
which quantities change is the same at all radii. In the case of a
centrifugally-supported disc, the exponential scale length is
determined (in a fixed potential) purely by the average angular
momentum per unit mass for the entire disc. Such a system thus has the
remarkable property that material over a significant range in radii
(and with a correspondingly significant range in specific angular
momenta) is somehow `informed' about the average specific angular
momentum of the disc as a whole.  The key question is whether this
property is a result of the formation process of the disc, or somehow
acquired through subsequent evolution.

Disc formation models have moved on from the picture of smooth
dissipational collapse within a dark halo potential, and recent
numerical calculations have focused on the more complex formation
histories resulting from hierarchical growth (eg. Kauffmann, White \&
Guiderdoni 1993; Cole et al 1994; Navarro, Frenk \& White 1995).  A
notable outcome of such calculations is the strong redistribution of
angular momentum during collapse. Attention has so far mainly focused
on the net transfer of angular momentum from baryonic to dark matter
and the consequent implications for the overall sizes of galactic discs
(eg. Navarro \& Benz 1991; Navarro \& Steinmetz 1997; Weil, Eke \&
Efstathiou 1998). A related issue is, however, the redistribution of
angular momentum within the baryonic material and its effect on the
resulting disc profile. The large dynamic range in size scales that
have to be captured in such simulations, and the difficulties in
controlling the effective viscosity in SPH calculations that are close
to the limit of adequate resolution,  mean that it is not yet clear
whether discs that are exponential over a number of scale lengths are
indeed the expected outcome.

Observational evidence against a picture of exponential infall followed
by {\it in situ} star formation is provided by the radial profiles of
gas in disc galaxies. The gas surface density profiles are generally
observed to be much flatter then the stellar profiles.  In the inner
regions of disc galaxies, the gas is the minority component and
represents the residue following star formation: it is not difficult,
by appropriate adjustment of the star formation law, to generate gas
profiles within the optical disc that are compatible with observations
using an initially exponential gas profile (eg.  Boissier \& Prantzos
1999). In the outer regions, however, the gas is the dominant component
and its profile should presumably reflect its infall distribution.  It
seems highly suspicious, in the context of {\it in situ} models, that
the gas profiles at large radii are so different from the exponential
profiles that characterise the stars at smaller radii: if the stellar
exponential profile derives from exponential infall of gas, surely one
would expect to see that profile continued in the gas not yet
incorporated in stars?

All of these difficulties are circumvented by the simple model first
proposed by Lin and Pringle (1987a). These authors envisaged that gas
in the disc not only turns into stars (on a timescale $t_*$), but is
also subject to some form of effective viscosity that gives rise to
angular momentum redistribution and radial flows.  The dominant
physical process responsible for the viscosity is poorly constrained,
but could plausibly result from cloud-cloud collisions and/or
gravitational instabilities in the disc.  If viscous redistribution
occurs on a timescale $t_{\nu}$, then the resultant stellar profile is
exponential for a wide range of initially less centrally concentrated
gas profiles, provided only that $t_{\nu} \sim t_*$. This result can be
understood, in the context of the above discussion, inasmuch as viscous
processes in the disc (provided $t_*$ is not much less than $t_{\nu}$)
allow material at different specific angular momenta to communicate
with each other, and thus to `inform' themselves of the average
specific angular momentum of disc material, which in turn sets the disc
scale length. On the other hand, $t_*$ cannot be much greater than
$t_{\nu}$ because in that case, viscous evolution would run away too
far towards its ultimate endpoint (i.e. `all the mass at the origin,
all the angular momentum at infinity', Lynden-Bell and Pringle (1974))
to be compatible with the observed profiles of disc galaxies.
Fortunately it turns out that the generation of exponential profiles
over a number of scale lengths does not require particularly fine
tuning between $t_*$ and $t_{\nu}$, and that satisfactory fits are
obtained even if the ratio $t_*/t_{\nu}$ deviates from unity by half an
order of magnitude, or is a mild function of radius (Clarke 1989).

 The viscous model thus relieves cosmological models of the burden of
generating the `right' angular momentum profile (though, by the same
token, it also erases information about the profile on infall, and thus
weakens the role of gaseous and stellar disc profiles as cosmological
probes). It also predicts that the radial gas surface density
distribution beyond the optical radius should be rather flat, in
accordance with observations. The viscous model was able to satisfy the
various observational constraints available a decade ago (mainly
surface density profiles and chemical constraints imposed by the Milky
Way), although it was not manifestly superior to conventional {\it in
situ} models in this respect (Clarke 1989, Yoshii \& Sommer-Larsen 
1989).  Then, as now, the main drawback to viscous models is the lack
of a detailed theory linking $t_*$ to $t_{\nu}$, although it is not
difficult to invoke physical processes (e.g. self-gravitational
instability of the disc) that one might expect to simultaneously
promote star formation and angular momentum redistribution in the disc
(Lin \& Pringle 1987b; Olivier, Primack \& Blumenthal 1991).

A wealth of new observational data pertaining to the evolution of disc
galaxies has emerged in recent years from both high redshift galaxy
surveys -- which permit a direct study of how galactic properties
evolve with epoch -- as well as detailed studies of the fossil stellar
populations in our own Milky Way and other local systems. Given that
the general arguments in favour of viscous models appear at least as
strong now as they did a decade ago, it would seem timely to revisit
this model and investigate its predictions in the light of new and
forthcoming observations.  In this paper, we explore the structural
evolution and radial growth of stellar discs subject to simultaneous
star formation and viscous evolution by considering models in which the
initial gas disc is pre-assembled as well as models in which the gas
disc is assembled over an extended period due to the continued infall
of material from the protogalactic halo. In both cases, the gaseous and
stellar discs are naturally built up from the inside-out, though the
mechanism is  different in each case: in pre-assembled models, gas is
transferred to larger radii only as a result of viscous torques in the
disc, whereas in the case of continued infall, the dominant cause of
the inside-out growth is the increasing specific angular momentum of
infalling material at later times (whether or not viscous flows are
also important).

We do not address in this paper how well viscous models can be made to
reproduce in detail the present-day properties of the Milky Way and
other nearby spirals; for a recent discussion of this aspect see Thon
\& Meusinger (1998).  We also bypass the issue of how one sets up the
conditions in the collapsing gas that will reproduce the correct {\it
absolute} values of the disc scale lengths.  The resultant scale length
in viscous evolution models is set purely by the average specific
angular specific angular momentum of the gas infallen at a given time
hence the problem of forming disc systems with large scale lengths (as
discussed by Mo et al 1998) is one of cosmological initial conditions
and not one that viscous evolution can alleviate or affect in any
way.   In the present paper, we simply assume that the forming disc has
access to material of the requisite high angular momentum. We describe
in Section \ref{sec:model} the model and  free parameters and present
our results in Section \ref{sec:results}.  Section \ref{sec:disc}
discusses model constraints from existing and future observational data
on both local and distant spirals.

\section {Model Parameters}  
\label{sec:model}
We envisage a general scenario for disc galaxy formation in which
infalling material gradually builds up the disc over a time interval
$t=0$ to $t=t_{infall}$ and (simultaneous) star formation and viscous
redistribution occur over the interval $t=t_{sf}$ ($\le t_{infall}$) to
$t=t_{now}$ ($\ge t_{infall}$). The case $t_{sf} = t_{infall}$
corresponds to the case of star formation in a pre-assembled gas disc
whereas $t_{now} \sim t_{infall} >> t_{sf}$ corresponds to the extreme
infall case where gas is accreted  up until the present epoch.

\subsection {Infall Calculation}
\label{sec:infall}
We calculate the surface density profile with which material arrives in
the disc plane by assuming an idealised (spherically symmetric) density
distribution and angular velocity distribution for baryonic material in
the protogalaxy.  We then map this material down into the disc plane by
invoking detailed angular momentum conservation, neglecting both the
self-gravity of the baryonic material and the response of the halo to
the infalling baryons. We justify this rigid halo approximation on the
grounds that the initial distribution of baryons in the halo is poorly
known (and likely to be considerably more complex than the smooth
spherically symmetric distribution invoked here) and also because in
any case, the behaviour of the viscous evolution is only weakly
dependent on the shape of the initial distribution (as opposed to its
total mass and angular momentum). We calculate the gas infall rate (for
$t \le t_{infall}$) by requiring that all material infalling from
within a spherical radius $r_{inf}(t)$ has reached the disc plane by a
time $t$.  This infall radius is fixed by the requirement that the free
fall time within $r_{inf}$ is much less than the instantaneous cosmic
shear time ($H^{-1}$) - or, equivalently, that the mean enclosed
density is much larger ($\approx 200$) than the mean density of the
Universe at that epoch. As successive shells of matter are `released'
and allowed to fall in, we map them onto the disc plane (without
applying the (small) correction for the finite infall time of each
shell).  We do not explicitly consider the gas cooling time in our
model;  numerical simulations suggest that the cooling radius tracks
the virial radius for Milky Way-type galaxies hence that all gas within
$r_{inf}$ is indeed able to infall on a free fall time (Somerville \&
Primack 1999).  Our default halo model assumes that the baryons
initially follow an $r^{-2}$ density distribution (ie.  isothermal
sphere), so that for an Einstein-de Sitter cosmology, the total infall
rate onto the galaxy is {\it constant} in time for $t < t_{infall}$.
At any given radius however, the infall history is characterised by an
initial lag of variable length (corresponding to the time required for
a shell of high enough angular momentum to fall in), a peak and then a
roughly exponential decline.  This prescription thus produces a crude
on/off switch on the infall, so that, if $t_{infall}$ is comparable to
$t_{now}$, it implies considerably more infall at recent times  and at
large radii than is usually assumed in galactic infall models, where
the infall is {\it arbitrarily} assumed to decline exponentially in both time
and radius (eg.  Boissier \& Prantzos 1999; Chiappini et al 1997)
in order to mimic inside-out growth.  We
motivate the prescription employed here by the wish to explore the
response of viscous discs to extreme infall conditions.

Finally, in order to prescribe the surface density distribution
generated by each shell as it falls in, we must assign an initial
angular momentum distribution for the baryons in the halo.  The angular
momentum of the dark matter and baryons is presumably acquired through
tidal torques from neighbouring structures (Peebles 1969) and thus the
average angular  momentum of infalling material (whether the infall be
smooth or discrete) is expected to increase with time.  We assume the halo
baryons initially rotate rigidly on spheres with angular velocity
$\propto r^{-1}$, which implies that the mean value of the spin
parameter $\lambda$ for each shell is constant.  For a power law
angular velocity law, the surface density distribution generated by
each shell is self-similar. We scale the overall dimensions and
normalisation of the system to match the observed profile of the Milky
Way.  As stated before, our calculation thus sidesteps the difficulty
of how, given the usual assumptions about the value of $\lambda$ and
the size scale for collapse as a function of redshift, it is possible
to produce large, Milky Way-scale, discs  at redshifts greater than
$\sim 1$ (eg.  Mo et al 1998).

\begin{figure}
\psfig{file=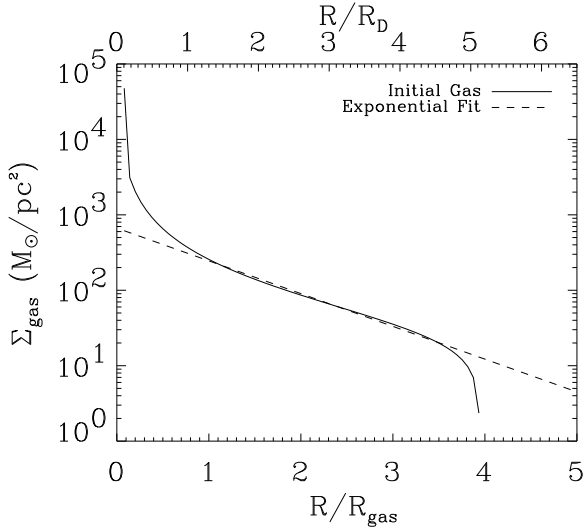,width=9cm}
\caption{The final gas surface density profile at the completion of the
infall phase (solid line) -- before star formation switches on --  for
a singular isothermal halo.   The bottom radial coordinate is
normalised to the scale length, R$_{gas}$ derived from an exponential
fit to the intermediate region of the profile (dashed line).  Note that
an exponential provides a good fit to the profile only over $\sim 2.5$
scalelengths.  The top radial coordinate is normalised to the scale
length of the stellar disc which ultimately forms from this gas
distribution, for the case of a star formation law $\propto
\Sigma_{gas}R^{-1}$. For the halo profile and rotation law adopted
here, $r_{init} \sim 5~R_D$.}
\label{fig:initial_infall}
\end{figure}

Figure \ref{fig:initial_infall} shows the final gas surface density
profile at the completion of the infall phase (hence the initial
starting point in the case of a pre-assembled gas disc).  Beyond the
central region, the gas surface density slowly declines and then turns
over abruptly at the maximum infall radius, $r_{init}$.  The value of
$r_{init}$/R$_D$ depends primarily on the initial halo profile and
rotation law, and to a lesser extent on the star formation and
viscosity prescriptions.  If the disc surface density decreases
outwards throughout the disc, $r_{init} > 3~R_D$; the absolute minimum
value of $r_{init}$, which corresponds to an annulus of matter at this
location, is $2~R_D$.  This latter configuration does not viscously
evolve to a smooth exponential profile however.  For the flat rotation
curve and halo profile assumed here (and neglecting disc self-gravity),
$r_{init} \sim 5~R_D$.   An exponential fit (dashed line)  provides a
good description of the radial behaviour of the initial gas profile
over $\sim 2.5$ scale lengths.

\subsection {Disc Evolution}
\label{sec:code}
 The evolution of gas surface density, $\Sigma_{gas}(r,t)$ in a disc
that is subject to simultaneous viscous redistribution and star
formation is described by:
\begin{equation}
\frac{\partial\Sigma_{gas}}{\partial{t}} = -\frac{1}{r} \frac{\partial}{\partial{r}}
\left\{\frac{\partial/\partial{r}[\nu\Sigma_{gas} r^3 (d\Omega/dr)]}{d/dr(r^2\Omega)}\right\}
-\frac{\Sigma_{gas}}{t_*}
\end{equation}
where $r, \Omega, \nu$ and $t_*$ represent the galactocentric radius,
angular velocity, kinematic viscosity and local star formation
timescale.  The evolution of the stellar surface density
$\Sigma_*(r,t)$ is then given by:
\begin{equation}
\frac{\partial\Sigma_*}{\partial{t}} = \frac{\Sigma_{gas}}{t_*}
\end{equation}

The gas surface density in these models is assumed to be the sum of the
neutral and molecular components.  We calculate the evolution of the
gaseous and stellar components of the disc, for times $t > t_{sf}$,
assuming that the gas evolves in the {\it fixed} potential of a halo
with flat rotation law (\ie $\Omega \propto r^{-1}$). This
simplification will not be correct for the very inner regions of the
disc, and so we do not attempt any detailed comparison of our results
with  structures in the central parts of disc galaxies.

In order to obtain a resulting exponential stellar profile, we follow
Lin \& Pringle (1987a) and fix the relationship between the star
formation and viscosity prescriptions such that the characteristic
timescale for material to move a fractional radius of order unity is
comparable with the timescale on which it is converted into stars
(i.e.  $t_* \sim t_\nu$). With this constraint, and given an assumed
rotation law, the only free parameter is the star formation
prescription. Our default  star formation law relates the star
formation rate per unit area with the gas column density,
$\Sigma_{gas}$, as $\Sigma_{gas}R^{-1}$, which corresponds to an $n=1$
Schmidt law, modulated by the local angular frequency  (eg. Wyse \&
Silk 1989).   A prescription of this sort (with power-law $n=1.5$) has
recently been shown to be in good agreement with the observed radial
star formation rate in the Milky Way (Boissier \& Prantzos 1999).   We
also consider a model with a dependence of $\Sigma_{SFR} \propto
\Sigma_{gas}^{1.5}$ as well as a model in which the star formation law
steepens consderably in the outer disc,  $\Sigma_{SFR}(R>3R_D) \propto
\Sigma_{gas}/R^2$.  While the former has been shown to provide a
reasonable description of the star formation rates observed in the
inner regions of galactic discs, the two component law  may provide a
better description of the steep declines in star formation rate
observed at large radii, perhaps resulting from the flaring of the gas
disc in these parts (eg. Ferguson et al 1998).  We  stress that in this
paper we are seeking generic properties of viscous models, and their
inter-relation with continued infall, that do not depend strongly on
the star formation prescription employed.  While varying the  star
formation prescription also requires changing the form of the viscosity
law so that the approximate equality in the respective timescales is
preserved, the dominant viscosity-producing mechanism in discs is so
poorly constrained that this does not create any obvious physical
problem\footnote{For a given star formation law of the form 
$\Sigma_{SFR} \propto \Sigma_{gas}^a R^{-b}$, the corresponding viscosity prescription is $\nu \propto \Sigma_{gas}^{a-1} R^{2-b}$}.

We compute the evolution of a viscous star forming disc subject to
infall using a standard first order explicit scheme with $200$
gridpoints, equispaced in $R$ with $R_{out}/R_{in} = 150$. Zero torque
boundary conditions are applied at the inner and outer edge of the
grid. In practice, we prescribe $t_{sf}$ and $t_{infall}$ and then pursue
the calculation until a time $t_{now}$ at which the system is deemed
to resemble the Milky Way - specifically, we require that the gas
fraction at a radius of $2.7$ exponential scale lengths\footnote{We
adopt R$_{\odot}=8.0$~kpc (Reid 1993) and R$_D$=3.0~kpc (Sackett
1997).} is $\approx 0.4$.

\section {Results} 
\label{sec:results}

\subsection{The Case of a Pre-Assembled Gas Disk}
\label{sec:preassevol}
\begin{figure*}
\centerline{\psfig{file=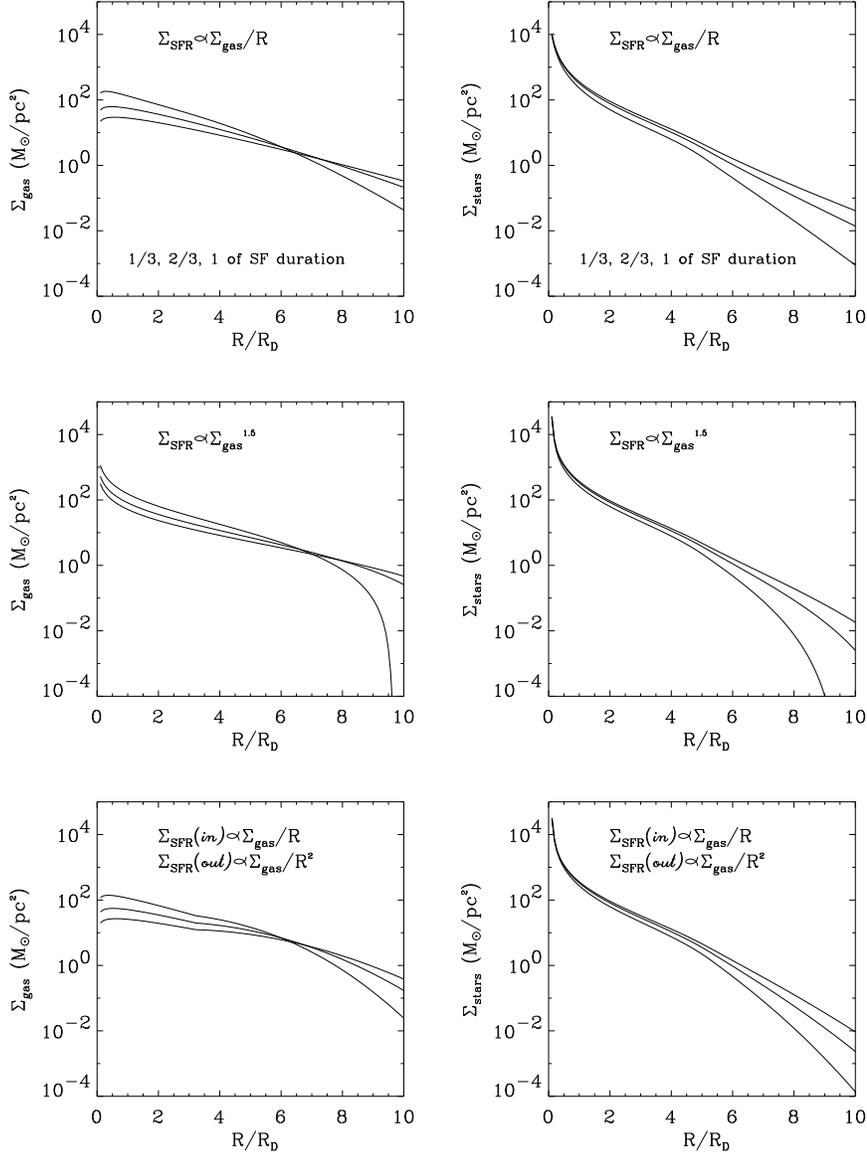,width=12cm}}
\caption{Snapshots of the gaseous (left) and stellar (right) surface
density distributions at various fractional times (1/3, 2/3 and 1) of
the star formation duration (which terminates at t$_{now}$) for the
case of a pre-assembled gas disc.  The radial coordinate is normalised
to the stellar exponential scale length measured at the end of the
model.  The final solutions are those with the highest stellar profile
at all radii and the lowest inner gas profile. 
The initial gas disc in these models extends to approximately
5 (final) exponential scale lengths.  Different star formation laws
are shown in top, middle and bottom panels.}
\label{fig:preassprof}
\end{figure*}

\begin{figure}
\psfig{file=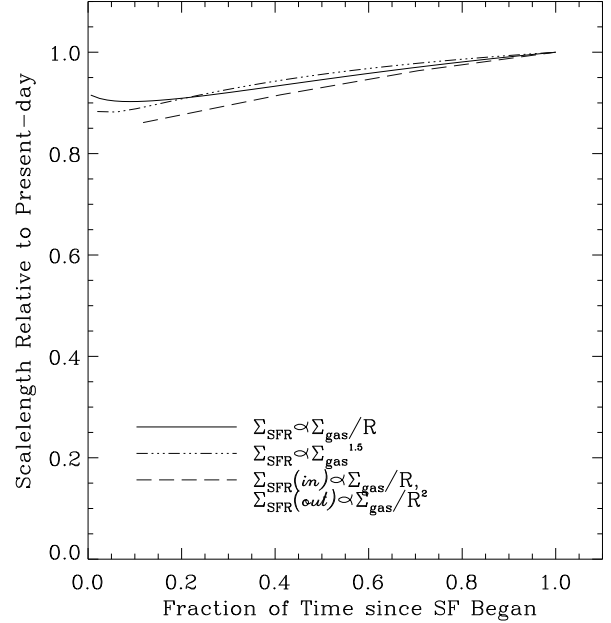,width=9cm}
\caption{Evolution of the stellar exponential scale length relative to
the present-day value as a function of time since the onset of star
formation for the case of a pre-assembled gas disc.  Results for
different star formation laws are shown. At each epoch, the stellar
scale length is measured by fitting the surface density profile over
the region corresponding to $\sim 0.3-0.9$ the radius of the initial gas
disc. }
\label{fig:preassevol}
\end{figure}

We first consider the results of models in which star formation (hence
viscous redistribution) switches on after the initial gas disc  is
already in place -- ie.  pre-assembled.  This corresponds to the
scenario where all or most of the infall phase is complete before the
onset of star formation ($t_{sf} \ge t_{infall}$).  Figure
\ref{fig:preassprof} shows snapshots of the gaseous and stellar surface
density distributions at various fractional times since the onset of
star formation for three different star formation prescriptions
($\Sigma_{SFR} \propto \Sigma_{gas}/R$, $\Sigma_{SFR} \propto
\Sigma_{gas}^{1.5}$ and a $`$two component' law described by
$\Sigma_{SFR} \propto \Sigma_{gas}/R^n$, where $n=1$ out to 3
scale lengths and $n=2$ beyond that).  The profiles are normalised by
matching the final (ie. present day) gas and stellar surface density in
the solar neighbourhood to their measured values (we assume
$\Sigma_{gas}=14.5~$M$_{\odot}$/pc$^2$ (Olling \& Merrifield 1998),
$\Sigma_{stars}=40~$M$_{\odot}$/pc$^2$ (Gould et al 1996)).   Figure
\ref{fig:preassprof} illustrates several notable features.  Firstly,
different star formation prescriptions lead to very similar exponential
behaviours (over at least 5 scale lengths) for the stellar surface
density profiles;  models really only deviate from each other at early
times and at large (unobservably faint) radii.  Interestingly, the two
component model produces rather smooth profiles despite the abrupt
change in the form of the star formation and viscosity laws at roughly
three scalelengths.  These particular profiles deviate somewhat from exponential
behaviour beyond 5--6 scale lengths, but the effect is subtle.   Gas
profiles are more sensitive to the different star formation laws, but
generally exhibit exponential declines in their outer regions with
e-folding scales significantly larger than that of the stars.   At late
times, the scale length of the gas at large radii, where the gas is
still the majority component, exceeds the stellar scale length by a
factor of approximately $2$.  Secondly,  we find the stellar component
exhibits self-similar growth out to $\sim 5$ scale lengths
over the entire star
forming period, implying an invariance in the stellar scale length
with time,  as first noted by Olivier et al (1991).  This is further
quantified  in Figure \ref{fig:preassevol} which shows that the scale
length lies well within $\sim 20\%$ of its present-day value throughout
the disc's history.  Thus, while viscous discs are characterised by
inside-out growth, which has led to the speculation that the resulting
scale length should increase with time (eg. Zhang \& Wyse 2000), we
here show that this growth is simply manifest as an expansion in the
radial extent of the exponential stellar disc due to the viscous
transport of gas from smaller radii.   This effect is unlikely to
be verifiable by studies of high redshift galaxies, however, since the
strongest evolution occurs in regions of the disc so faint that they
are inaccesible to direct observation.

\begin{figure}
\psfig{file=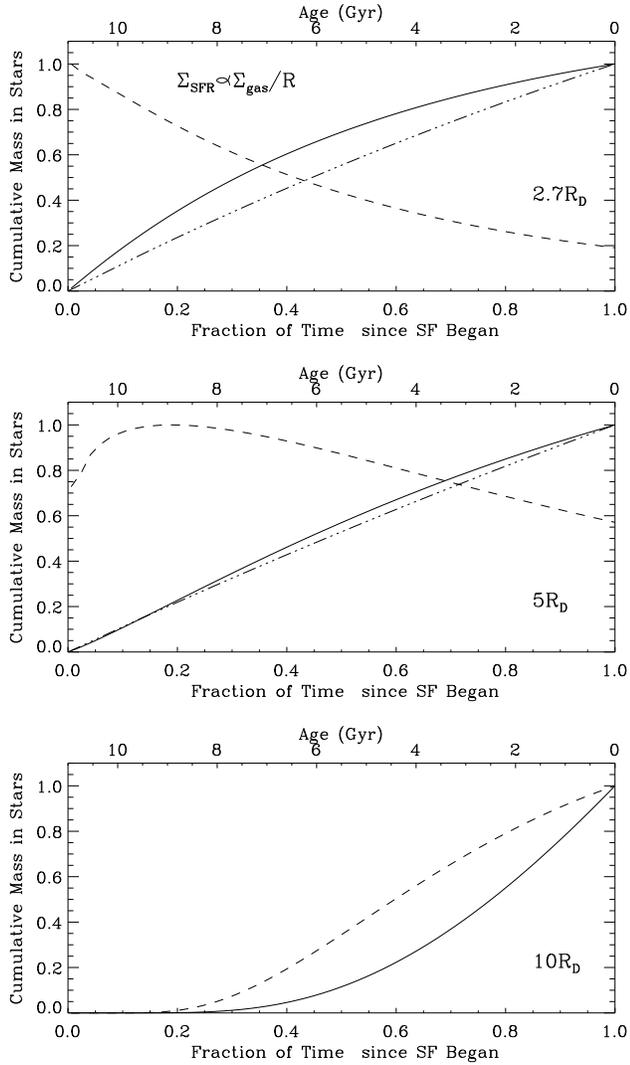,width=9cm}
\caption{The star formation history (dashed line) of the disc at radii
corresponding to  $2.7$ (top), $5$ (middle) and $10$ (bottom)
exponential scale lengths in the case of a pre-assembled gas disc.
The star formation rate is normalised to the peak rate at each radius.
These models are for a star formation law
$\Sigma_{SFR}=\Sigma_{gas}/R$.  The top axis shows the age of the disc,
assuming the present-day age of 11~Gyr (Binney et al 2000).  Also shown
is the fraction of the final stellar mass in place (solid line) as a
function of time and the corresponding analytic prediction of an {\it
in situ} model (dashed-dotted line) obeying the same star formation
law.  We do not show the {\it in situ} predictions for the disc at 10
scale lengths, since this lies well beyond the extent of the initial
gas disc where gas would never reach in the absence of viscous flows. }
\label{fig:preassages}
\end{figure}

Figure \ref{fig:preassages} shows the star formation rate (dashed
lines) normalised to the peak rate as a function of time  at  $2.7$
(`solar neighbourhood'), $5$ and $10$ exponential scale lengths for the
particular star formation law $\Sigma_{SFR} \propto \Sigma_{gas}/R$;
other star formation prescriptions exhibit similar behaviours.  For a
Freeman exponential disc ($\Sigma_B = 21.65$ mag/$\Box^{''}$), 5 scale
lengths corresponds to an expected surface brightness of $\sim 1\%$ of
sky, or about the current limit for quantitative studies of surface
brightness in external galaxies.  The top axis shows age of the disc in
Gyr, where we have assumed the present-day stellar disc age of the
Milky Way to be 11~Gyr  as determined from the ages of the oldest stars
in the Hipparcos sample (Binney, Dehnen \& Bertelli 2000). The disc at
2.7 and 5 scale lengths exhibits slowly declining star formation rates
in the pre-assembled model, decreasing by factors of roughly $5$ and
$2$ over the disc's lifetime respectively.  On the other hand, the disc
at 10 exponential scalelengths is characterised by a slowly increasing
star formation rate (bottom panel Figure \ref{fig:preassages})
reflecting the finite time required to transport gas outwards to these
parts by viscous torques; the arrival of material in the extreme
outermost disc is therefore delayed long after the onset of star
formation at smaller radii.   This delay is not a major effect at $5$
scale lengths however, even for the most compact plausible initial gas
distributions.

Also shown in Figure \ref{fig:preassages} is the fraction of the final
stellar mass in place as a function of time (solid lines).  These
curves are contrasted with the analytic predictions of a simple  {\it
in situ} model (dashed-dotted line) obeying the same star formation law
(note that in the {\it in situ} model the total - gas $+$ stars -
radial profile is exponential by construction at all times) calculated
as:
\begin{equation}
\Sigma_{SFR}=\frac{\partial{\Sigma_{gas}}}{\partial{t}} =
\frac{k\Sigma_{gas}}{R} \end{equation} where the adopted star formation
efficiency, $k$, is chosen to be consistent with the viscous model.
Integration yields \begin{equation}
\frac{\Sigma_{stars}(t)}{\Sigma_{stars}(t_{now})} =
\frac{1-e^{-kt/R}}{1-e^{-kt_{now}/R}} 
\end{equation} 
where we have normalised the mass in stars present at a given time to
that of the present day.  While the mean age of the stellar population
present at $5$ scale lengths is somewhat smaller than at $2.7$ scale
lengths, the effect is rather small and the bulk of the stars at this
radius have formed fairly early on.  Specifically, while 50\% of the
final stellar surface density is in place after $0.3$ of the star
formation duration (or $\sim$ 3.5~Gyr for our assumed disc age) at
$2.7$ scale lengths, this same amount is in place after $\sim  0.4$ of
the star formation period (or $\sim$ 5~Gyr) at $5$ scale lengths.
Interestingly, there does not appear to be any great difference between
the growth rate of stellar mass at $5$ scale lengths  in the viscous
and {\it in situ} models, implying that viscous flows have little
impact on the star formation history at these radii; observations of
stellar age distributions alone in these parts could not distinguish
between viscous and {\it in situ} scenarios.  Beyond the radius of
the initial gas disc, the stellar population becomes 
increasingly young with radius.

\subsection{The Case of Extreme Infall}
\label{sec:infallevol}
\begin{figure*}
\centerline{
\psfig{file=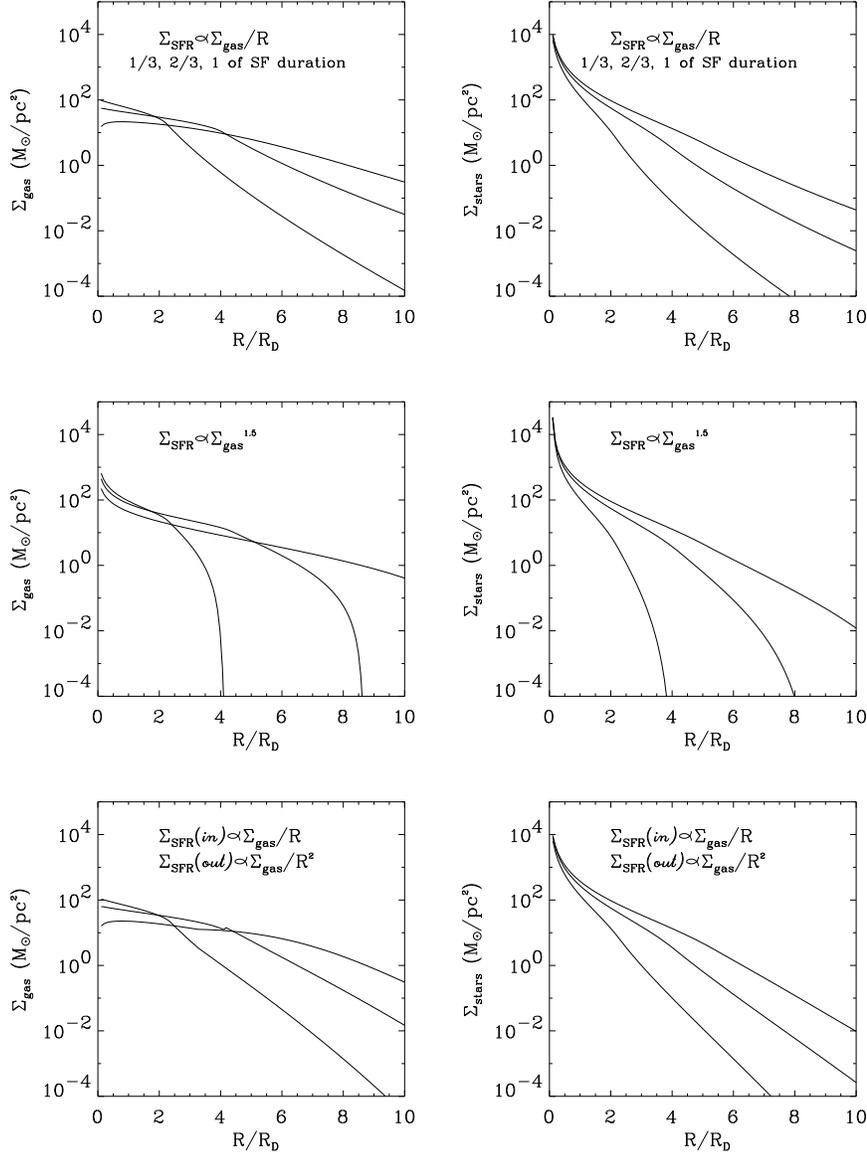,width=12cm}}
\caption{Snapshots of the gaseous (left) and stellar (right) surface
density distributions at various fractional times (1/3, 2/3 and 1) of
the star formation duration (which terminates at t$_{now}$) for the
extreme infall model.  In this model star formation starts shortly
after the infall phase begins ($t_{star} \sim \frac{1}{10}t_{infall}$.
The radial coordinate is normalised to the stellar exponential scale
length measured at the end of the model.  The final solutions are those
with the highest stellar profile at all radii and the lowest inner gas
profile.  Different star formation
laws are shown in top, middle and bottom panels.}
\label{fig:infallprof}
\end{figure*}

\begin{figure}
\psfig{file=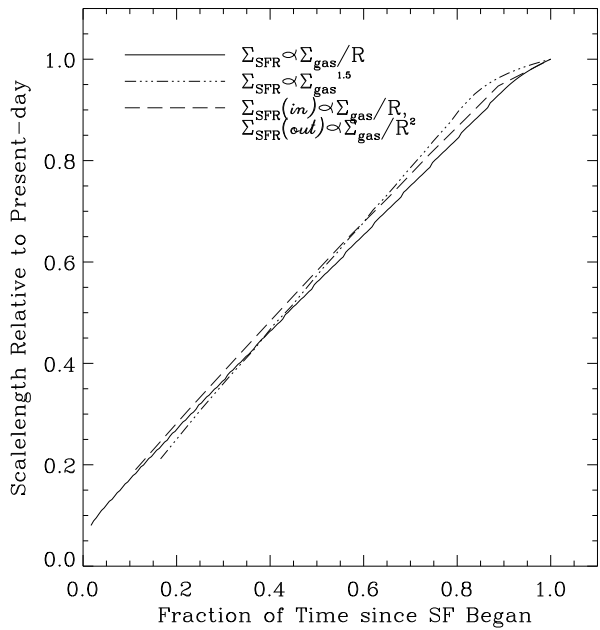,height=9cm}
\caption{Evolution of the stellar exponential scale length relative to
the present-day value as a function of the star formation duration for
the constant infall model.  Results for different star
formation laws are shown. At each epoch, the stellar scale length is
measured by fitting the surface density profile over the region
corresponding to 0.3--0.9 of the instantaneous infall radius.  }
\label{fig:infallevol1}
\end{figure}

We now contrast the models above with viscous models in which $t_{infall}
\sim t_{now}$.  We refer to this as  the $`$extreme infall' case, since
 the total mass infall rate onto the galaxy has been constant up to
nearly the present day, and  star formation in the disc started 
early on in the infall period ($t_{star} \sim \frac{1}{10} t_{infall}$).
Figure \ref{fig:infallprof} shows the stellar and gaseous surface
density profiles for different star formation prescriptions at various
times since the onset of star formation.  For all models, the final
stellar and gaseous profiles are almost identical for the extreme
infall viscous model and the pre-assembled viscous model (contrast the
highest stellar profiles and lowest central gas profiles in Figures
\ref{fig:preassprof} and \ref{fig:infallprof}).  The temporal evolution
of these profiles is very different however.  While the disc is able to
continuously adjust to the changing profile of the infallen material so
as to generate an approximately exponential stellar profile at all
times, the scale length of the exponential evolves systematically with
time, simply reflecting the mean specific angular momentum of the
material which has fallen in by that epoch. The stellar profiles are
always exponential over at least 4--5 scale lengths.   For the density
and angular momentum distributions in the halo assumed here ($\rho
\propto r^{-2},  \Omega \propto r^{-1}$), the total angular momentum
scales with the square of the infallen mass, so that the mean specific
angular momentum is simply proportional to the infallen mass.  Our
default infall prescription, which makes specific assumptions about the
halo properties and the cosmological model (see Sec \ref{sec:infall}),
implies a constant infall rate for $t \le t_{infall}$ and hence the mean
specific angular momentum of the infallen material rises linearly with
time over this period. Figure \ref{fig:infallevol1} demonstrates that
the stellar scale length indeed displays this behaviour.

\begin{figure}
\psfig{file=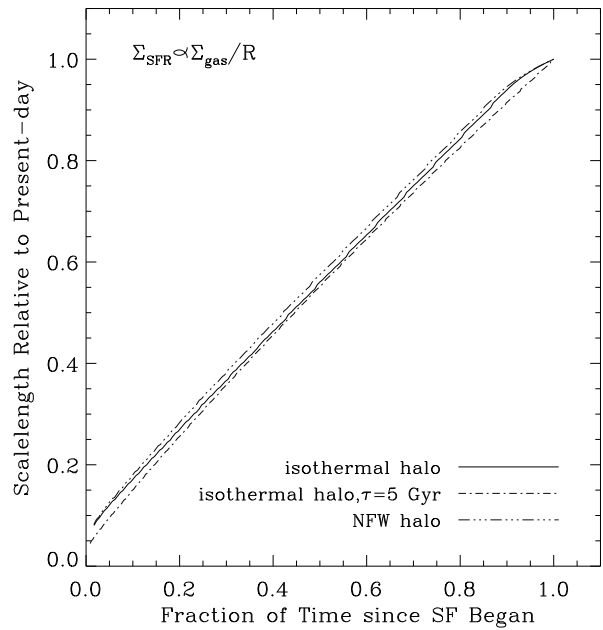,height=9cm}
\caption{Evolution of the stellar exponential scale length relative to
the present-day value as a function of different infall prescriptions
(ie. halo profiles).
As in Figure \ref{fig:infallevol1}, infall has been occuring in these
models since the onset of star formation.   At each epoch, the stellar scale length is
measured by fitting the surface density profile over the region
corresponding to 0.3--0.9 of the instantaneous infall radius.   }
\label{fig:infallevol2}
\end{figure}

 A fundamental assumption in our infall model is that the mean specific
angular momentum of the accreted material systematically increases with
time.  We feel this is a valid assumption, regardless of the exact
nature of the infall, and should be preserved. For example, if the
dominant mode of infall is high velocity cloud accretion (Blitz et al
1999), one would still expect those clouds falling in at late times to
have originated further from the Milky Way and to have suffered {\it on
average} more tidal torquing than clouds which fell in at early times.
While individual clouds will carry a range of angular momenta (and
indeed in our models a given shell maps onto a range of radii), the
mean trend should be for the average angular momentum to increase with
time.  It is worthwhile investigating how much of the scale length
evolution is due to other aspects of our adopted infall model however.
Figure \ref{fig:infallevol2} shows the growth of the stellar
exponential scale length for an isothermal halo model in which the
total infall rate is allowed to  decrease exponentially with time.  In
this case, the amount of recent infall (ie. t $\sim$ t$_{now}$) is
reduced relative to that in the constant infall case by a factor
$\approx$ 11~Gyr/$\tau_{inf}$, where 11~Gyr is the present-age of the
disc and $\tau_{inf}$ the e-folding time of the infall rate.  In the
limit of $\tau_{inf} >> 11$~Gyr, this infall prescription mimics the
constant infall case whereas  for $\tau_{inf} << 11$~Gyr it mimics the
pre-assembled disc case, with essentially all the infall occuring early
on; Figure \ref{fig:infallevol2} shows an intermediate case where
$\tau_{inf}=$5~Gyr. We also examine an infall model in which the
baryons in the halo are described by a Navarro-Frenk-White (NFW)
profile in which $\rho \propto (r/r_s)^{-1}(1+r/r_s)^{-2}$ with the
parameter r$_s$ defining the scale radius.  Navarro, Frenk \& White
(1997)  argue on the basis of high resolution N-body simulations that
this form better characterizes the equilibrium  density profiles of
dark matter halos than an isothermal sphere.  In terms of infall rates,
it leads to a somewhat higher infall rate than the isothermal sphere at
early times and somewhat lower at late times.  Figure
\ref{fig:infallevol2}  indicates that the growth of the scale length in
these alternate infall models is very similar to that in the constant
infall case and thus that our results on the scale length evolution are
robust.  Contrasting Figures \ref{fig:infallevol1} and
\ref{fig:infallevol2} with Figure \ref{fig:preassevol}, one concludes
that if clear evidence were to emerge for an increase in stellar scale
length of galaxies over cosmological time, it would, within the
framework of our model, be {\it an unambiguous indicator of  the late
infall of high angular momentum material} and could not merely be
attributed to the effects of viscous evolution.

\begin{figure}
\psfig{file=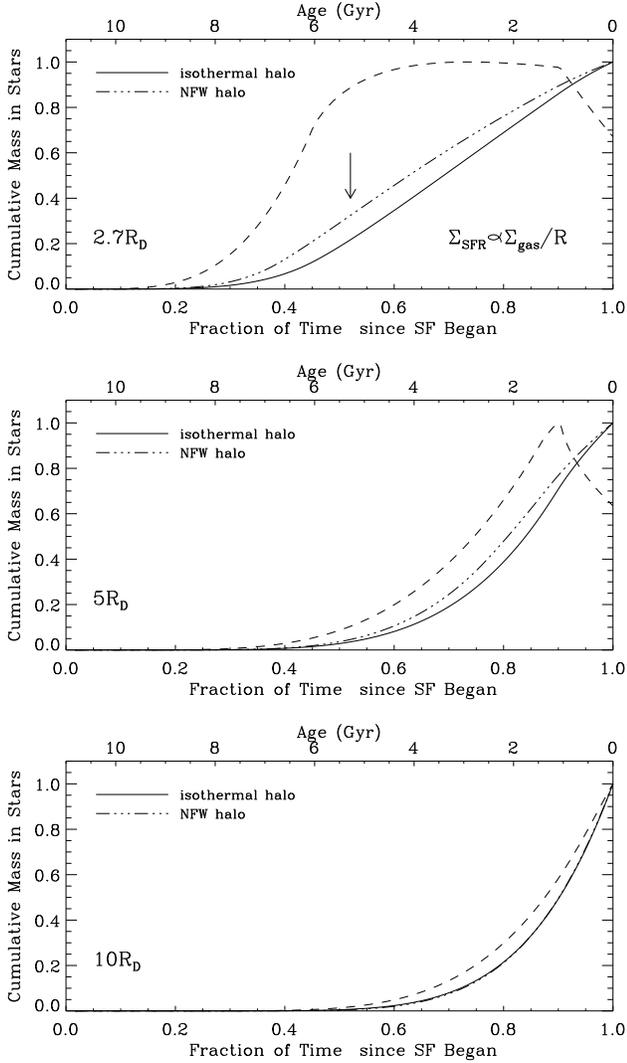,width=9cm}
\caption{The star formation history (dashed line) of the disc at radii
corresponding to  $2.7$ (top), $5$ (middle) and $10$ (bottom)
exponential scale lengths in the extreme infall case.  The star
formation rate is normalised to the peak rate at each radius.  These
models are for a star formation law $\Sigma_{SFR}=\Sigma_{gas}/R$.  The
top axis shows the age of the disc, assuming the present-day age of
11~Gyr (Binney et al 2000).  Also shown is the fraction of the final
stellar mass in place for the particular cases of baryonic infall from
halos with isothermal and NFW forms (solid and dashed-dotted lines
respectively) as a function of time.  The time corresponding to the
peak infall rate at the solar neighbourhood is marked by the arrow in
the top panel.}
\label{fig:infallages}
\end{figure}

A consistent picture emerges from analysis of the star formation
history at different locations in the disc.  Figure
\ref{fig:infallages} shows the star formation rate (dashed line)
normalised to the peak rate as a function of time for our default star
formation law.   Very few stars form initially at the solar
neighbourhood as a direct result of the low gas density (only gas
viscously transported from smaller radii is present).  As the peak
infall rate at this radius is reached (indicated by the arrow in the
top panel of Figure \ref{fig:infallages}), the star formation rate
rapidly increases and then levels off to an almost constant value until
the present epoch.  The fraction of the final stellar mass in place as
a function of time for singular isothermal sphere and NFW halo baryon
distributions is also shown (solid and dashed-dotted lines
respectively).    As expected, the disc at $2.7$ scale lengths is
predominantly young in the extreme infall model, with only a small
dependence on the specific form of the infall prescription.
Specifically, 50\% of the final stellar mass in place after 0.7 of the
star forming duration, or $\sim 8$~Gyr if a disc age of 11~Gyr is
assumed (note that in this case very few 11~Gyr stars are expected to
exist outside the innermost regions of the disc, where the first infall
is received).   At $5$ and 10 scale lengths, the stellar population is
conspicuously young in the extreme infall model.  The disc at 5 scale
lengths receives material from direct infall of halo gas only at the
very end of the infall phase, so that prior to this, gas only reaches
this location as a result of outward viscous transport from an
initially much more compact gas distribution.  This is especially true
of the disc at 10 scale lengths, which  never receives direct infall in
our model.  Consequently, the dominant star-forming phase in these
regions only gets under way at late times.  We draw attention to the
fact that this combined infall/viscous model may produce an unusual
chemical signature in the very outer regions, comprising an older metal
rich population (formed from pre-enriched gas transported outwards from
small radii at early times) and a young metal poor population
containing stars formed from recent unenriched infall.  We plan to
investigate detailed predictions for the chemical signatures of viscous
evolution and infall in a future paper.

\section{Discussion}
\label{sec:disc}
We have explored models for the evolution of disc galaxies subject to
star formation, viscous redistribution of angular momentum,  and
cosmologically-motivated gaseous infall.   We have focused attention on
the extreme cases of these models and the rather different predictions
which arise for the temporal evolution of the exponential scale length,
and the radial growth rate of galactic stellar discs.  In this section,
we discuss how these model predictions compare with relevant
observational data for both local and distant disc galaxies.

\subsection{Constraints from the Local Universe}
Stellar populations in the Milky Way and other nearby spirals provide the
$`$fossil record' of star formation and galaxy assembly over cosmic
lifetimes.  The wealth of high quality data available for stars in our
local neighbourhood of the  Milky Way disc makes it a logical starting
point for model comparisons.

An important prediction of our models is the star formation history --
which is reflected in the stellar age distribution -- at various
radii.  The pre-assembled and extreme infall models make very different
predictions for the star formation history of the disc at  2.7
exponential scale lengths.   In the former case, the star formation
rate declines smoothly by a factor of $\sim 5$ over the disc's lifetime
with the consequence that a significant fraction of disc stars form
early on.   In the latter case, however, very few stars form initially
with the rate then increasing steeply to a roughly constant value.
 As a consequence, most stars form late in this model.  Quantitatively,
for an assumed  disc age of 11~Gyr, more than 50\% of the final stellar
mass is in place at solar neighbourhood after $\sim 3.5$~Gyr in the
pre-assembled model whereas less than 5\% is in place by same time in
extreme infall model.

\begin{figure}
\psfig{file=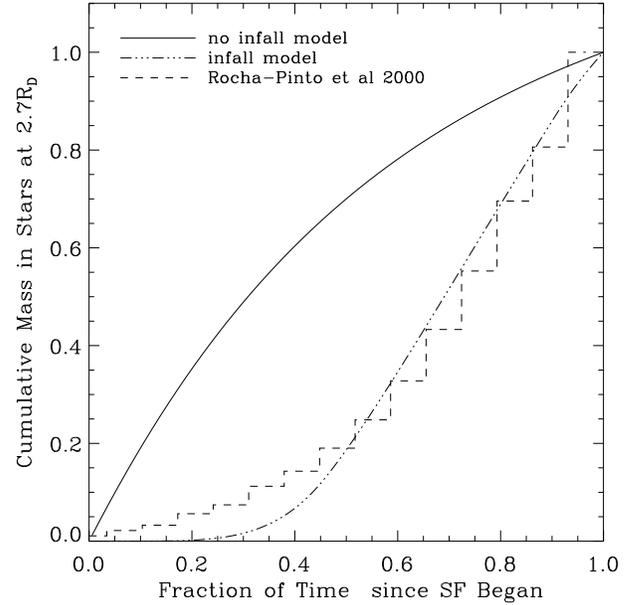,width=9cm}
\caption{The cumulative mass in stars as a function of time in the
solar neighbourhood as derived from the chromospheric age distribution
of Rocha-Pinto et al (2000) (dashed line). Also shown are the predictions
for our extreme infall model (dashed-dotted line) and the pre-assembled disc
model (solid line). }
\label{fig:rpcomp}
\end{figure}

It is well-established that stars with a considerable range in ages
exist in our local neighbourhood, a non-negligible fraction of which
appear to be $`$old', ie. ages $\gtrsim 8$~Gyr (eg. Edvardsson et al
1993, Binney et al 2000, Rocha-Pinto et al 2000).   Rocha-Pinto et al
(2000) have recently re-determined the star formation history  of the
local Galactic disk from a sample of 552 late-type dwarfs exhibiting
chromospheric activity. Stellar ages are derived using a new
metallicity-dependent chromospheric activity-age relation (see their
Table 3). Assuming this sample provides an unbiased representation of
the total thin disc population, we use their age distribution to infer
the cumulative mass in stars present in the solar neighbourhood as a
function of time as shown in Figure \ref{fig:rpcomp}.  This curve shows
a  good agreement with the predictions of our extreme
infall viscous model.  While the Rocha-Pinto data suggest a slightly
higher star formation rate at early times (hence a small excess of very
old stars), the overall similarity of the two star formation histories
at later times is quite striking.   We reiterate that our star formation
history is largely determined by the infall history at this radius, and
not by viscous flows.  Although the possibility remains that some
observational biases exist within the Rocha-Pinto et al sample -- for
example, the onset and duration of stellar chromospheric activity is
poorly understood (eg.  Balnius \& Vaughan 1985) --  it seems unlikely
that these biases could alter the derived star formation history enough
to make it more closely resemble the predictions of our pre-assembled
model.

Knowledge of the age distributions of stars at large radii in discs
(for example, at 5 scale lengths) would have even greater leverage for
model comparisons, since it is precisely these parts where the
predictions of the extreme infall and pre-assembled models deviate the
most.  Unfortunately, essentially no information is currently available
on the range of stellar ages present in these parts.  Due to our
location within the Milky Way disc, this issue may be more easily
addressed in the future by deep studies of the resolved stellar
populations in nearby external galaxies than in our own.

The effects of angular momentum transport alone mean that viscous
models naturally predict inside-out growth for galactic discs whether
or not infall is also present.   Associating the $`$mean age' of the
stellar population  with the age by which 50\% of the final stellar
mass is in place, our models suggest an age gradient of $\sim -0.7$~Gyr
over the radial range spanning  2--5 scale lengths in discs (see
Figures \ref{fig:preassages} and \ref{fig:infallages}).  Inside-out
formation has long been suspected in disc galaxies on the basis of
radial colour and metallicity gradients (eg.  Tinsley \& Larson 1978).
For example, de Jong (1996b) finds that radial colour gradients (in the
sense of bluer colours at larger radii) extending over several scale
lengths are common in spirals and are largely consistent with trends in
stellar age alone.   Using stellar population synthesis models, Bell
and de Jong (2000) have explicitly quantified the radial age variation
implied by these colour gradients and find the typical large spiral to
exhibit an age gradient of $\sim -0.8$~Gyr per K-band scalelength,
which also agrees with the predictions of our models.

Finally, we note that there is still significant debate as to the
importance of continued gaseous infall for disc galaxy evolution, with
few existing direct observational constraints.   Galactic high velocity
clouds may be a signature of gas accretion onto the Milky Way, but the
lack of information concerning the distances of these clouds hinders a
clear understanding of their origin.  Blitz et al (1999) have carried
out a simple simulation of the formation and evolution of the Local
Group in which 10$^6$ particles (each representing a 10$^6$~\msun\ gas
cloud) are subject to the gravitational forces of M31, the Milky Way
and the tidal field of the other Local Group members.  Any cloud
falling towards the Milky Way or M31 which passes within 100 comoving
kpc of their centres is assumed to be accreted by that galaxy.  They
find present-day neutral gas accretion rates of $\sim 1$\msun~yr$^{-1}$
for both galaxies, with the rates being somewhat higher in the distant
past. The overall evolution is consistent with an exponential decline
with an e-folding time of about $5 \times 10^9$\yr.  On the other hand,
our extreme infall model, when normalised to the Milky Way, requires a
constant infall rate of $4.5$\msun~yr$^{-1}$.  Our infall models
therefore require gas accretion rates that are slightly higher but of
the same order of magnitude as those found in Blitz's simulation.

\subsection{Constraints from Distant Galaxies}
An exciting recent development has been the availability of
quantitative structural parameters for moderate redshift disc galaxies
which can be used to directly map the size evolution of galaxies with
cosmic epoch (eg. Roche et al 1998, Lilly et al 1998, Simard et al
1999, Giallongo et al 2000).  Based on a sample of discs drawn from the
CFRS and LDSS redshift surveys, Lilly et al (1998) find the size
function of disc scale lengths in disc-dominated galaxies (ie.
bulge-to-total ratios $\leq 0.5$) to be roughly constant to z$ \sim 1$,
at least for large discs with R$_D > 3.2 h^{-1}_{50}$~kpc.  In a
similar study,  Simard et al (1999) analyse the size-magnitude relation
defined by disc-dominated field galaxies in the DEEP survey.  Carefully
considering possible selection effects, these authors also conclude
that there is no significant evidence for any evolution in the
size-magnitude distribution of disc galaxies over the redshift range
$0.1 < z < 1.1$ and that a significant number of discs lie close to the
canonical Freeman relation at all redshifts probed.

Taken together, these studies suggest only a very modest change in
stellar exponential scale length with epoch out to $z \sim$ 1 (however
see Mao et al (1998) for an alternative interpretation of the same
dataset).   For an Einstein de Sitter Universe,  the interval between a
redshift of unity and the present epoch can be associated with  the
period from $\sim 0.3-1.0$ times the star formation interval in our
models, assuming that star formation in the disc begins early on.
Inspection of Figures \ref{fig:infallevol1} and \ref{fig:infallevol2}
indicates that the disc scale lengths are expected to increase by about
a factor of 3 over this redshift range if gaseous infall is important.
Thus, in contrast to the Milky Way constraints discussed in the
previous section, it would appear that observations of moderate
redshift galaxies lend support for a scenario in which continued infall
-- as we have modelled it -- is {\sl not} important for building
galactic discs.  We note that the global  star formation rates of the
discs which form in our pre-assembled and extreme infall models
decrease and remain roughly constant with time respectively, neither of
which are inconsistent with current observations of large disc
galaxies.
  
It should be kept in mind, though, that quantitative characterization
of distant galaxy profiles is still plagued by a number of
difficulties.  For the highest redshift galaxies in the aforementioned
samples, exponential profiles are generally fit over 1--2 disc scale
lengths and hence are particularly susceptible to  errors in the bulge
and disc separation.  Furthermore, studies to date  have measured
structural parameters in a single passband over the redshift range of
interest. If strong  colour gradients are present in these systems,
this could conspire with the $k$-correction to mimic a constant scale
length in a fixed passband over a range in redshift even though the
rest-frame scale lengths are evolving.   Colour gradients significantly
in excess of those measured in local galaxies would be required for
this to be an important effect however.

\section {Conclusions}
We have explored models for the evolution of disc galaxies that invoke
simultaneous star formation and viscous redistribution of angular
momentum, as well as a cosmologically-motivated prescription for
gaseous infall.    The existence of galaxies which exhibit exponential
disc profiles over significantly  more than $\sim 5$ scale lengths (eg.
Weiner et al 2000, Barton \& Thompson 1997) lend strong support for
viscous models as viscosity may be the only viable mechanism for
transporting material to these extreme radii, while maintaining a
smooth exponential profile in the stellar disc.   Following  Lin \&
Pringle (1987a), we fix the relationship between the star formation and
viscosity prescriptions such that the characteristic timescale for
material to move a fractional radius of order unity is comparable with
the timescale on which it is converted into stars (i.e.  $t_* \sim
t_\nu$).  While the inclusion of viscous flows is essential for
ensuring that the stellar disc is always exponential, we find that such
flows play essentially no role in determining the evolution of the disc
scale length, or the star formation history over the inner $\sim 5$
scale lengths (corresponding to the maximum infall radius in our
model).  Indeed,  the dominant process governing the size evolution and
star formation history of the luminous stellar disc is that of gaseous
infall.
  
We summarize our main results as follows:\\
1. In models in which the main infall phase precedes the onset of star
formation (ie. a pre-assembled gas disc), we find the exponential scale
length  to be rather {\it invariant} with time over the entire history
of the disc.  This result holds true for a variety of popular star
formation prescriptions.  The value of the stellar disc scale length is
primarily determined by mean specific angular momentum of the initial
gas disc.     \\
2.  In the pre-assembled model, the star formation rate smoothly
declines with time within the inner  5 scale lengths (approximately the
maximum infall radius, $r_{init}$).  As a result, there is a
considerable $`$old' (ie. $> 8$~Gyr) population present at these
radii,  with half of the final stellar mass  in place more than 6~Gyr
ago.  The onset of star formation is delayed at radii greater than
$r_{init}$ as the outer disc is formed entirely by material viscously
transported from smaller radii; the stellar populations in these parts
are thus increasingly young with radius.\\
3. If infall continues during the main star-forming phase of the
galaxy's evolution, the stellar disc remains approximately exponential
at all times, but now the scale length {\it increases towards the
present epoch}, in simple proportion to the average specific angular
momentum of the material fallen in to the galactic disc to date. For
example, for an isothermal halo density profile and constant spin
parameter, the scale length is directly proportional to the infallen
mass which is proportional to time.  Regardless of whether viscosity is
at work or not, late infall of high angular momentum material is the
only way to increase the stellar scale length of discs within the
framework of our models.
4. In contrast to the case of a pre-assembled disc, the disc stellar
populations in the infall models are predominantly young, with more
than half of the stars even at the location of solar neighbourhood
predicted to  have  ages $\lesssim 5$~Gyr.  The disc at 5 scale lengths
receives direct infall of halo gas only at the very end of the
accretion phase, so that prior to this, gas only reaches this location
as a result of outward viscous transport.  As in the pre-assembled
case, all material at radii beyond the maximum infall radius has been
viscously transported there from smaller radii.\\
5. We compare our model predictions with recent observations of disc
galaxies sizes and stellar populations.  A particularly good agreement
is found between the solar neighbourhood star formation history
predicted by our extreme infall model and the recent observational
determination of this quantity by Rocha-Pinto et al (2000).  On the
other hand, the relative invariance of disc scale length with cosmic
epoch, as derived from studies of moderate redshift galaxies, appears
to lend support for our pre-assembled model.   Future observations that
will be important for clarifying this situation include detailed
studies of resolved stellar populations at large galactocentric radii
($\gtrsim$ 5 scale lengths) in the Milky Way and other nearby spirals
in order to constrain the age and star formation history of the outer
disc, and larger, deeper surveys of the structural parameters  of
moderate redshift disc systems in various passbands.

\section{Acknowledgments}
We are grateful to Dave Schade, Nicole Vogt, Rosie Wyse and especially Rachel
Somerville for useful discussions during the course of this work.

\end{document}